\documentclass[twocolumn,american,prl,superscriptaddress]{revtex4}
\usepackage[T1]{fontenc}
\usepackage[latin1]{inputenc}
\usepackage{amsmath}
\usepackage{graphicx}
\usepackage{amssymb}

\makeatletter
\usepackage{ae,aecompl}

\usepackage{babel}
\makeatother

\begin{document}

\title{Supersolid behavior of nonlinear light}

\author{Albert Ferrando}
\affiliation{Departament
d'\`Optica, Universitat de Val\`encia. Dr. Moliner, 50. E-46100 Burjassot
(Val\`encia), Spain.}
\altaffiliation{Interdisciplinary Modeling Group, InterTech (http://www.intertech.upv.es)}
\author{Miguel-\'Angel Garc\'{\i}a-March}
\affiliation{Instituto
Universitario de Matem\'atica Pura y Aplicada (IUMPA), Universidad Polit\'ecnica
de Valencia. Camino de Vera s/n. E-46022 Valencia, Spain.}
\altaffiliation{Interdisciplinary Modeling Group, InterTech (http://www.intertech.upv.es)}
\author{Mario Zacar\'es}
\affiliation{Instituto
Universitario de Matem\'atica Pura y Aplicada (IUMPA), Universidad Polit\'ecnica
de Valencia. Camino de Vera s/n. E-46022 Valencia, Spain.}
\altaffiliation{Interdisciplinary Modeling Group, InterTech (http://www.intertech.upv.es)}

\date{\today{}}

\begin{abstract}

We present a formal demonstration that light can simultaneously exhibit
a superfluid behavior and spatial long-range order when propagating
in a photonic crystal with self-focussing nonlinearity. In this way,
light presents the distinguishing features of matter in a {}``supersolid''
phase. We show that this supersolid phase provides the stability conditions
for nonlinear Bloch waves and, at the same time, permits the existence
of topological solitons or defects for the envelope of these waves.
We use a condensed matter analysis instead of a standard nonlinear
optics approach and provide numerical evidence of these theoretical
findings.
\end{abstract}


\maketitle

Matter can undergo a phase transition at ultracold temperatures in
which exhibits a superfluid behavior and, at the same time, present
all the characteristics of a crystalline solid. This new phase of
matter is known as {}``supersolid'' and it has been experimentally
proven in helium-4 in recent years \citep{kim-nature427_225}. Apparently,
light is unrelated to these phases characteristic of condensed matter.
However, analogies between condensed matter and optical systems are
increasingly appearing in the literature \citep{freedman-nature440_7088}.
Even the concept of liquid phase of light has been already suggested
\citep{michinel-pre65_066604}. In this letter, we will take this
analogy a step further and we will demonstrate a formal equivalence
between regular light structures propagating in a nonlinear photonic
crystal and ultracold matter in a {}``supersolid'' phase. For the
purpose of establishing this equivalence, we will use the formalism
of condensed matter and particle physics \citep{chaikin00}
instead of using a standard nonlinear optics approach.

In Hamiltonian formalism the nonlinear Schr\"odinger equation for a
periodic medium with Kerr nonlinearity can be obtained from the Hamiltonian
density as $i\partial\phi/\partial z=\partial\mathcal{H}/\partial\phi^{*}=\left(-\nabla_{t}^{2}+V(\mathbf{x})-g|\phi|^{2}\right)\phi$.
In our case $\nabla_{t}$ is the 2D transverse gradient operator and
the potential $V(\mathbf{x})=-\left(n^{2}\left(\mathbf{x}\right)-n_{0}^{2}\right)$
represents the periodic modulation of the square of the refractive
index with respect to the reference value $n_{0}^{2}$. Transverse
and axial coordinates are normalized. The energy of a propagating
solution is given by $E=\int d^{2}x\mathcal{H}(\phi).$\textbf{\emph{\large{}
}}Now, instead of using $E$ we introduce the optical equivalent of
the free energy in statistical physics $F=E-\mu P=\int d^{2}x\mathcal{F}(\phi)$
where $\mu$ is the propagation constant of a stationary solution
and $P$ is the system power (whose, in absence of losses, is a constant
in the propagation). Thus $P$ plays the role of $N$, the particle
number in statistical physics, and $\mu$ plays the role of the chemical
potential. In terms of the free energy density the equation of motion
is $i\partial\bar{\phi}/\partial z=\partial\mathcal{F}/\partial\bar{\phi}^{*}=\left(-\nabla_{t}^{2}+V(\mathbf{x})-\mu-g|\bar{\phi}|^{2}\right)\bar{\phi}$.
It is easy to check that the relation between the solutions of both
equations is simply $\phi=\bar{\phi}e^{-i\mu z}$. Thus an stationary
solution $\phi$ with propagation constant $\mu$ is equivalent to
a $z$-independent solution $\bar{\phi}$ that is an extremum of $\mathcal{F}$
verifying $\partial\mathcal{F}/\partial\bar{\phi}^{*}=0$. For the
case under consideration, the optical free energy would be given by
(we use $\phi$ instead of $\bar{\phi}$):\begin{equation}
F=\int d^{2}x\left[\nabla_{t}\phi^{*}\nabla_{t}\phi+V(\mathbf{x})|\phi|^{2}-\mu|\phi|^{2}-\frac{g}{2}|\phi|^{4}\right].\label{eq:free_energy}\end{equation}
In this context the typical nonlinear optics $P(\mu)$ curve for a
stationay solution appears as the conventional equation of state of
statistical physics $P=-\partial F/\partial\mu$.

On the other hand, we are interested in the analysis of stationary
solutions whose amplitude is invariant under finite translations of
value \textbf{$\mathbf{a},$} where $\mathbf{a}$ is the period of
the potential. If $\phi_{\mathrm{sol}}$ is a solution that satisfies
the equation for stationary states and that simultaneously verifies
the condition $|\phi_{\mathrm{sol}}(\mathbf{x}+\mathbf{a})|=|\phi_{\mathrm{sol}}(\mathbf{x})|$
then the total potential $V_{\mathrm{sol}}(\mathbf{x})\equiv V(\mathbf{x})-g|\phi_{\mathrm{sol}}(\mathbf{x})|^{2}$
occurring in such equation will be periodic with period given by $\mathbf{a}$,
$V_{\mathrm{sol}}(\mathbf{x}+\mathbf{a})=V_{\mathrm{sol}}(\mathbf{x})$.
 Self-consistency implies that $\phi_{\mathrm{sol}}$ must be equal
to one of the Bloch functions $\phi_{\mathrm{sol}}(\mathbf{x})=f_{\mathbf{Q};\beta_{0}}(\mathbf{x})=e^{i\mathbf{Q}\cdot\mathbf{x}}v_{\mathbf{Q};\beta_{0}}(\mathbf{x})$
forming the spectrum of the nonlinear operator generated by itself
and it will be characterized by a propagation constant $\mu=\mu_{\mathbf{Q};\beta_{0}}$.
Using the nonlinear operator generated by $\phi_{\mathrm{sol}}$ (including
the total potential $V_{\mathrm{sol}}(\mathbf{x})\equiv V(\mathbf{x})-g|\phi_{\mathrm{sol}}(\mathbf{x})|^{2}$)
we construct a \emph{nonlinear }Wannier function basis by means of
standard techniques. We assume that all functions are defined within
a periodic region $\Omega$, called the basic domain, of spatial dimensions
$Na\times Na$ so that we represent any arbitrary field amplitude
at a given axial position $z$ as $\phi(\mathbf{x},z)=\sum_{\hat{i},\alpha}c_{\hat{i},\alpha}(z)W_{\alpha}^{\mu}(\mathbf{x}-\mathbf{x}_{\hat{i}}).$The
nonlinear Bloch solution $\phi_{\mathrm{sol}}(\mathbf{x})$ is represented
by the site-independent and $z$-independent coefficients $c_{\hat{i}}^{\mathrm{sol}}=\eta_{\mu}e^{i\mathbf{Q}\cdot\mathbf{x}_{\hat{i}}}$
where $\eta_{\mu}\equiv\sqrt{P}/N$. Note that $\mathbf{Q}$ is discretized
because $\phi$ is defined in the periodic domain $\Omega$ so that
$\mathbf{Q_{m}}=2\pi\mathbf{m}/(Na)$ where $\mathbf{m}\in\mathbb{Z}^{2}$
and fulfills $-N/2\le m_{x},m_{y}\le N/2$. We introduce now the Wannier
expansion into the expression for the optical free energy and integrate
out the tranverse coordinates $\mathbf{x}$ to obtain \begin{eqnarray}
F & = & -t(\mu)\sum_{\hat{i}}\sum_{\nu=1}^{2}c_{\hat{i}}^{*}\left(c_{\hat{i}+\hat{n}_{\nu}}+c_{\hat{i}-\hat{n}_{\nu}}\right)+\nonumber \\
 &  & +l(\mu)\sum_{\hat{i}}|c_{\hat{i}}|^{2}-\frac{U(\mu)}{2}\sum_{\hat{i}}|c_{\hat{i}}|^{4}+\left(\mathrm{h.o.t}.\right),\label{eq:free_energy_c}\end{eqnarray}
where $\hat{n}_{1}$ and $\hat{n}_{2}$ are the {}``horizontal''
and {}``vertical'' lattice vectors, $t(\mu)\equiv-L_{\hat{0},\hat{0}+\hat{n}_{1}}=-L_{\hat{0},\hat{0}+\hat{n}_{2}}$
is the effective nearest neighbors coupling and $l(\mu)\equiv L_{\hat{0}\hat{0}}$
and $U(\mu)\equiv2T_{\hat{0}\hat{0}\hat{0}\hat{0}}$ are the two on-site
couplings. The second and fourth order couplings $L_{\hat{i}\hat{j}}(\mu)$
and $T_{\hat{i}\hat{j}\hat{k}\hat{l}}(\mu)$ are obtained as overlapping
integrals of the nonlinear Wannier basis associated to the stationary
solution and, for this reason, they depend on $(\mu,\mathbf{Q},\beta_{0})$
(for simplicity, we disregard interband interaction terms and elliminate
band indices): $L_{\hat{i}\hat{j}}(\mu)\equiv\int_{\Omega}d^{2}xW_{\hat{i}}^{\mu*}\left(-\nabla_{t}^{2}+V(\mathbf{x})-\mu\right)W_{\hat{j}}^{\mu}$
and $T_{\hat{i}\hat{j}\hat{k}\hat{l}}(\mu)\equiv\frac{1}{2}\int_{\Omega}d^{2}xgW_{\hat{i}}^{\mu*}W_{\hat{j}}^{\mu*}W_{\hat{k}}^{\mu}W_{\hat{l}}^{\mu}.$
The expression $(\mathrm{h.o.t}.)$ in \eqref{eq:free_energy_c} stands
for higher-order terms involving interactions at longer distances.

The main difference between \eqref{eq:free_energy_c} and previous
approaches is that the nonlinear Wannier basis provide different coefficients
than those obtained using linear Wannier functions or localized single
potential solutions as in the tight binding aproximation. Since our
aim is to analize the stability behavior of the nonlinear Bloch solution
characterized by $\mu$ the election of the nonlinear Wannier basis
associated to it is a natural choice. With this purpose in mind let
us write the $z$-dependent Wannier coefficients as $c_{i}(z)=e^{i\mathbf{Q}\cdot\mathbf{x}_{\hat{i}}}\Phi_{i}(z)=e^{i\mathbf{Q}\cdot\mathbf{x}_{\hat{i}}}\left(\eta_{\mu}+\Delta\Phi_{i}(z)\right)$
to formalize the fact that we want to analyze the dynamics associated
to the stationary solution described by $c_{i}^{\mathrm{sol}}=\eta_{\mu}e^{i\mathbf{Q}\cdot\mathbf{x}_{\hat{i}}}$.
In this way the dynamic information is encoded in the envelope coefficients
$\Phi_{i}(z)$. 

In this Letter we are interested in the regime where perturbations
have a correlation length larger than the potential period ---$\zeta\gg a$.
Physical phenomena in this regime present a collective spatial character
that will reflected in the properties of the discrete envelope function
$\Phi_{i}(z)$. Spatial collective effects will be represented by
values of $\Phi_{i}$ that will fluctuate smoothly in space. It is
then natural to take the continuous limit of the discrete optical
free energy by considering the limit $a/\zeta\rightarrow0$ and by
introducing the continuous spatial envelope function $\Phi(\mathbf{x}_{t},z)$
defined as $\Phi(\mathbf{x}_{\hat{i}},z)=\Phi_{i}(z)$. We substitute
$c_{i}=e^{i\mathbf{Q}\cdot\mathbf{x}_{\hat{i}}}\Phi_{i}$ into \eqref{eq:free_energy_c}
and then take the continuous limit using standard techniques. In order
to simplify the result, we consider only {}``diagonal'' nonlinear
Bloch waves with $Q_{1}=Q_{2}=2\pi m/(Na)$, $m\in\mathbb{Z}$. The
phase difference between two neighboring sites of the nonlinear Bloch
wave is then $\Delta\phi_{m}=2\pi m/N$.  In order to relate the
optical effects driven by $F$ to condensed matter and particle physics
phenomena we need to re-define the optical free energy in some cases.
For this reason, we introduce the sign-changed optical free energy
defined as $\bar{F}\equiv\mathrm{sign}(b_{m})F$

\begin{figure}
\includegraphics[width=8cm]{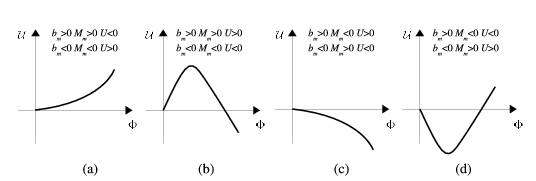}

\caption{Effective potential for different configurations
\label{fig:Effective-potential}}

\end{figure}

\begin{eqnarray}
\bar{F} & = & \int_{\Omega}d^{2}x\left[|b_{m}(\mu)|\nabla\Phi^{*}\nabla\Phi+i\mathrm{sign}(b_{m})(\mathbf{v}_{m}(\mu)\cdot\nabla\Phi)\Phi^{*}\right.\nonumber \\
 &  & \left.+\mathcal{U}(|\Phi|^{2})+\left(\mathrm{h.o.t}.\right)\right],\label{eq:sign_free_energy}\end{eqnarray}
where $b_{m}(\mu)\equiv\cos(\Delta\phi_{m})t(\mu)a^{2},$ $\mathbf{v}_{m}(\mu)=v_{m}(1,1)$
with $v_{m}(\mu)\equiv-2\sin(\Delta\phi_{m})t(\mu)a$ and $M_{m}(\mu)\equiv l(\mu)-4t(\mu)\cos(\Delta\phi_{m}).$
In this case $\left(\mathrm{h.o.t}.\right)$ includes terms with higher-order
derivatives and $\mathcal{U}=\mathrm{sign}(b_{m})\left(M_{m}(\mu)|\Phi|^{2}-\frac{U(\mu)}{2}|\Phi|^{4}\right)$.
In principle, the use of $F$ or $\bar{F}$ is irrelevant as far as
dynamics is concerned since dynamics remains the same under the change
$F\rightarrow-F$. The advantage in the use of $\bar{F}$ is that
it contains \emph{in all cases }a positive definite kinetic energy
term, as required in quantum field theory to define a proper vacuum
state \citep{coleman85}. By doing this, equivalences are straightforward
to achieve even with negative values of $b_{m}$. By means of $\bar{F}$
we can proceed to determine the nature of the ground state of the
system by analyzing the behavior of the so-called effective potential
$U(\Phi)$ {}``\`a la Landau''. Landau theory is a mean-field approach
used to characterize phase transitions in condensed matter and particle
physics \citep{chaikin00}. The qualitative form of the
effective potential is represented in Fig. \eqref{fig:Effective-potential}
for different signs of coefficients. One can easily recognize in this
figure that there are only two configurations for which a spatially
uniform envelope solution $\Phi(\mathbf{x})=\Phi_{0}$ is allowed
(cases (b) and (d)). They correspond to extrema of the optical free
energy density $\partial\mathcal{F}/\partial\Phi^{*}=\partial\mathcal{U}/\partial\Phi^{*}=0$
given by the condition $\left|\Phi_{0}\right|=M_{m}(\mu)/U(\mu)$.
The ground state of the system is the state that minimizes the free
energy, thus, only the two cases in (d) can provide an spatially homogeneous
ground state. The former analysis is identical to that performed in
condensed matter physics to establish the nature of phase transitions
in the mean-field regime, here $\mu$ playing the role of temperature.
In this way, the signs of the $b_{m}(\mu)$, $M_{m}(\mu)$ and $U(\mu)$
coefficients determine the nature of the ground state and, therefore,
in which\emph{{}``phase'' }light is. It is important to strees that
the $\mu$ dependence in these {}``Landau coefficients'' is the
result of using nonlinear Wannier functions. The ground state in Fig.\ref{fig:Effective-potential}(d)
is degenerate since all solutions of the type $\Phi_{\phi}=|\Phi_{0}|e^{i\phi}$
are minima of the effective potential and thus they all have the same
free energy. The optical free energy is invariant under a $U(1)$
phase transformation $\Phi\rightarrow e^{i\alpha}\Phi$. The ground
state $\Phi_{\phi}$, however, is not since this phase transformation
maps it into a different degenerate solution $\Phi_{\phi+\alpha}$
with the same free energy. This mechanism is well-known in condensed
matter and particle physics and it is known as \emph{spontaneous symmetry
breaking} (SSB). Superfluidity, superconductivity, the Higgs mechanism
or the chiral phase transition in quantum chromodynamics are physical
phenomena related to the same mechanism. The SSB mechanism have distinctive
properties: (i) appearence of a non-zero order parameter in the broken
phase, (ii) existence of topological solitons or defects, (iii) presence
of masless or long-range excitations (Goldstone bosons). In this language,
our optical system in the Fig.\ref{fig:Effective-potential}(d) configuration
is in a \emph{broken phase}, a phase in which $U(1)$ symmetry has
been spontaneously broken. We expect then to find the optical counterparts
of the aforementioned properties. In this Letter we will pay attention
to properties (i) and (ii), leaving the analysis of (iii) for a further
publication.

We can now establish a link between the stability of nonlinear Bloch
waves and the SSB mechanism. A nonlinear Bloch wave is characterized
by an homogeneous envelope. According to our previous analysis, only
in the configurations in Fig.\ref{fig:Effective-potential}(d) this
envelope function corresponds to the ground state of the optical free
energy. On the other hand, in this configuration the system is in
the $U(1)$ broken phase. Since the ground state is a stable solution,
i.e., fluctuations around it cannot transform this solution into a
different one by any dynamical mechanism, we infer that the stability
condition of a nonlinear Bloch wave is achieved in the broken phase,
i.e., when ($b_{m}>0$, $M_{m}<0$, $U<0$) or ($b_{m}<0$, $M_{m}>0$,
$U>0$). The properties of a stable nonlinear Bloch wave are then
identical to that of a superfluid as far as its envelope is concerned.
However, the solution simultaneously presents spatial long-range order.
Using quantum field theory notation, if $\left|0\right\rangle $ stands
for the ground state given by the nonlinear Bloch wave in the mean-field
regime at a given $\mu$: (i) $T_{\mathbf{a}}\left|0\right\rangle =e^{i\mathbf{Q}\cdot\mathbf{a}}\left|0\right\rangle $
where $T_{\mathbf{a}}$ is the lattice translation operator and (ii)
$\left\langle 0\right|\hat{\Phi}\left|0\right\rangle =|\Phi_{0}|e^{i\phi}\neq0$.
Physically speaking, the {}``soliton lattice'' filling the periodic
medium completely and described by the nonlinear Bloch wave presents
perfect spatial long-range order as in a cristallyne solid and, at
the same time, it possesses a non-vanishing order parameter indicating
it is in a superfluid phase. The dynamics of low energy fluctuations
around such a solution is determined by Eq.\eqref{eq:sign_free_energy}
and has to show identical features than a superfluid. In this way,
we demonstrate that, under the specified conditions, light fulfills
the definition of a supersolid, that is, it is a spatially ordered
system (like in a solid or crystal) with superfluid properties. %
\begin{figure}
\includegraphics[height=3cm]{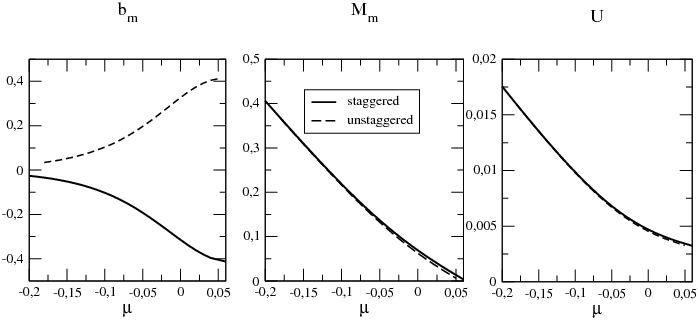}

\caption{Landau coefficients ($b_{m}$, $M_{m}$, $U$) in terms of the propagation
constant $\mu$ for unstaggered (dashed line) and staggered (solid
line) nonlinear Bloch solutions.\label{fig:Landau-coefficients}}

\end{figure}
\begin{figure}
\begin{tabular}{cc}
 (a)&
(b)\tabularnewline
\includegraphics[%
  scale=0.2]{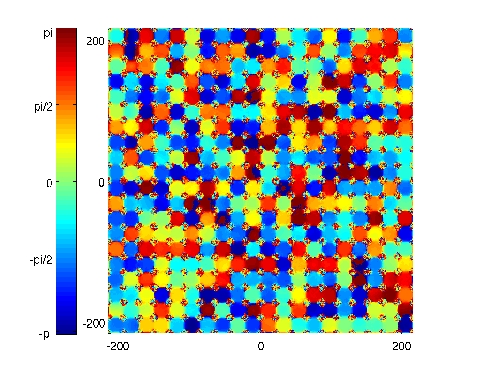} &
\includegraphics[%
  scale=0.2]{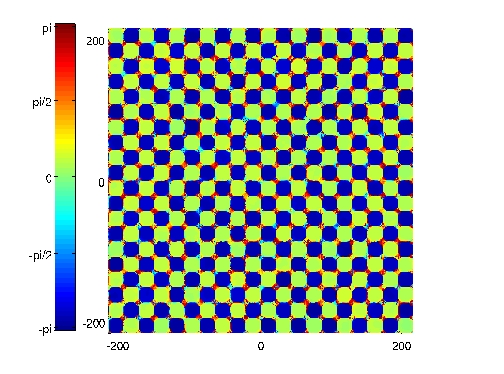} \tabularnewline
\end{tabular}
\caption{Stability/unstability patterns of phase under propagation:(a) for
a perturbed unstaggered solution phase disorders quickly, (b) for
a staggered solution spatial order remains. \label{fig:Stability/unstability}}

\end{figure}
\begin{figure}
\begin{tabular}{ccc}
 (a)&
(b)&(c)\tabularnewline

\includegraphics[%
  scale=0.32]{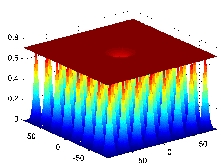}&
\includegraphics[%
  scale=0.32]{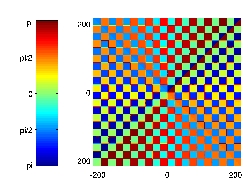}&
\includegraphics[%
  scale=0.32]{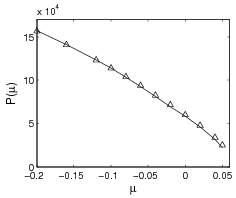}
 \tabularnewline

\end{tabular}

\caption{Supersolid light topological defect for $\mu=-0.1$, $V_{0}=2$, and charge $+1$: (a) simultaneous representation of the amplitudes
of the envelope $\Phi$ and full $\phi$ functions for this solution,
(b) representation of its phase and (c) $P(\mu)$ diagram calculated
using the envelope equation (solid line) and the full equation (triangles).
\label{fig:topological-defect}}

\end{figure}

We have performed a number of numerical experiments to check both
qualitative and quantitative the validity of our condensed matter
approach. We have modelled a nonlinear photonic crystal with \emph{$\mathcal{C}_{4}$
}symmetry formed by a nonlinear material with refractive index $n_{1}$
with embedded circular inclusions of a linear material with index
$n_{2}<n_{1}$, the index contrast being given by the potential $V_{0}=-\left(n_{2}^{2}-n_{1}^{2}\right)>0$.
This type of nonlinear photonic crystal can be achieved in standard
or in chalcogenide photonic crystal fibers \citep{ferrando-oe11_452}
or in laser-writing waveguides \citep{blomer-oe14_2151}.
The nonlinearity is self-focussing and of the Kerr type $V_{\mathrm{NL}}=-g|\phi|^{2}$
($g>0$). First of all, we have evaluated the dependence of the {}``Landau
coefficients'' ($b_{m}$, $M_{m}$, $U$) on $\mu$ . We have proceeded
as follows: (i) we numerically evaluate a nonlinear Bloch solution
$\phi_{\mathrm{sol}}$ at a given $\mu$, (ii) we determine the nonlinear
operator associated to $\phi_{\mathrm{sol}}$ (i.e., including the
nonlinear potential $V_{\mathrm{NL}}=-g|\phi_{\mathrm{sol}}|^{2}$)
which numerically will be a matrix, (iii) we find the spectrum of
this matrix formed by Bloch eigemodes since the total potential, including
the linear and nonlinear part, is periodic, (iv) we construct the
\emph{nonlinear} Wannier basis out of these Bloch modes using the
standard technique \citep{ashcroft76} and (v) we evaluate the {}``Landau
coefficients'' using their definitions as overlapping integrals of
Wannier functions. A given nonlinear Bloch wave characterized by $\mu$
then univocally provides a specific value for ($b_{m}$, $M_{m}$,
$U$). In Fig. \ref{fig:Landau-coefficients} we represent the functional
form of these coefficients in terms of $\mu$ for two solutions with
different pseudo-momentum \textbf{$\mathbf{Q}$}: a solution with
$\mathbf{Q}=0$ with a phase difference between neighboring sites
$\Delta\phi_{m}=0$ (unstaggered) and a solution with $\mathbf{Q}=(\pi/a,\pi/a)$
with $\Delta\phi_{m}=\pi$ (staggered). By analyzing the signs of
their Landau coefficients we immediately recognize these two configurations
correspond to the cases (b) (unstaggered) and (d) (staggered) in Fig.\ref{fig:Effective-potential}.
The main difference between these two configurations arise from the
different sign of $b_{m}$, which has its origin in their different
phase structure since $b_{m}\sim t(\mu)\cos(\Delta\phi_{m})$. In
the case analyzed here $t(\mu)>0$ for all $\mu$ in both configurations,
so that the different form of the effective potential in both cases
is a pure effect of the underlying phase of the nonlinear Bloch wave.
According to our condensed matter analysis, the unstaggered solution
can exist as an homogeneous solution but it corresponds to a metastable
state that eventually will decay into a different state. It cannot
be stable. On the contrary, the staggered solution is the ground state
of a light supersolid and, consequently, it is necessary stable. We
have performed a numerical stability analysis of these two configurations
that confirm these predictions. A small initial perturbation of the
unstaggered solution gives rise, after a short propagation, to the
breaking of spatial long-range order crearly reflected in the progressive
disordering of the phase of the solution ---see Fig.\ref{fig:Stability/unstability}(a).
The staggered solution is, however, immune to perturbations and preserve
spatial long-range order in phase and amplitude ---see Fig.\ref{fig:Stability/unstability}(b).
Similar examples can be found in the literature that can be explained
by the same mechanism, e.g., stability of large truncated nonlinear
Bloch waves \citep{petrovic-pre68_55601}.
As mentioned before, one of the most paradigmatic properties of a
superfluid is the existence of topological solitons or defects. They
are holes in the superfluid around which the superfluid flows with
a quantized circulation. Therefore, it is expected that the envelope
function $\Phi$ supports the existence of the optical counterparts
of these objects in the broken phase. We have proven indeed the existence
of these type of solutions by numerically solving the same nonlinear
photonic crystal structure using: (i) the \emph{effective} equation
for $\Phi$ associated to $\bar{F}$ ---Eq.\eqref{eq:sign_free_energy}---
using the values of ($b_{m}$, $M_{m}$, $U$) given in Fig.\eqref{fig:Landau-coefficients}
and (ii) the full equation for the original field $\phi$ including
the periodic potential $V(\mathbf{x})$. It is remarkable that both
approaches are in excellent agreement with each other both qualitatively
and quantitatively. As correctly predicted by theory, the topological
vortex only exists \emph{on top of }the staggered solution ---see
Fig.\ref{fig:topological-defect}(b)--- which corresponds to the ground
state in the broken superfluid phase $|\Phi_{0}|\neq0$. From an optical
point of view, one could consider this object as a type of dark soliton,
however, the striking property of this dark soliton is that exists
in a \emph{self-focusing} medium when they are usually associated
to defocusing media. In Fig.\ref{fig:topological-defect}(a) we can
appreciate the excellent fit of the envelope function to the solution
of the full equation. This excellent quantitative agreement is even
more clearly appreciated in the calculation of the $P(\mu)$ curve
for a family of topological vortices using both approaches shown in
Fig.\ref{fig:topological-defect}(c). In summary, Figs.\ref{fig:topological-defect}(a)
and (b) show the simultaneous presence of superfluid behavior (nontrivial
amplitude and phase of the topological defect) and spatial long-range
order (perfect staggered order of the background in all the domain).
Besides, numerical stability analysis indicates that these solutions
are stable under propagation during long distances. These features
represent a clear numerical evidence of the supersolid behavior of
light in nonlinear photonic crystals predicted by theory. In this
sense, it has been shown that light can be treated analogously as
matter not only in their qualitative aspects but by using a condensed
matter formalism. We call this approach to nonlinear optics \textit{photonic
condensed matter}.

This work was partially supported by the Government of Spain (contracts
FIS2005-01189 and TIN2006-12890) and Generalitat Valenciana (contract
APOSTD/2007/052).

\end{document}